\documentclass[11pt]{article}
\usepackage{hyperref}
\begin{document}
\title{Surface Texture and Pulsation Due to  \\ Balloon Bursting in Different Liquids}
\author{Enrique Soto and Andrew Belmonte \\
\\\vspace{6pt} The W.G. Pritchard Laboratories, Department of Mathematics \\ Penn State University, University Park, PA 16802, USA}
\maketitle
\begin{abstract}

We study the instabilities occurring during the burst of an air
balloon in a liquid. These instabilities are typical for the
deformation of an interface between two fluids of different
densities, similar to fingering in Rayleigh-Taylor instability
(see e.g. Sharp, 1984). In
\href{http://hdl.handle.net/1813/11466}{Video} a series of bursts
are shown for air balloons in different liquids. When the balloon
tears it tracks the surface, generating wrinkles and releasing the
pressure inside. Apparently, the texture of the surface during the
burst becomes smoother as the viscosity increases. During the
burst the surface breaks and generates several small bubbles.
Furthermore, the pressure inside the balloon is higher than the
external pressure before the burst; once the balloon tears the
pressure is released and the generated bubbles pulsate several
times (see e.g. Brennen, 1995). Such oscillations are more evident
for higher internal pressures.
\end{abstract}
\begin{enumerate}
\item Sharp, D.H. \emph{An Overview of Rayleigh-Taylor
Instability}, Physica , \textbf{12D}, 3-18 (1984).

\item Brennen, C.E \emph{Cavitation and Bubble Dynamics}, Oxford
University Press, Chap \textbf{3} (1995).

\end{enumerate}

\end{document}